\documentclass{llncs}
\usepackage{graphicx}
\usepackage{amstext}
\usepackage{amssymb,amsmath}
\usepackage{subfigure}
\usepackage{algorithm}
\usepackage{algorithmic}

\begin{document}

\title{Detecting a Corrupted Area \\ in a 2-Dimensional Space}

\author{YounSun Cho}

\institute{Purdue University\\
\email{cho52@purdue.edu} 
}

\maketitle

\begin{abstract}
\medskip
Motivated by the fact that 2-dimensional data have become popularly used 
in many applications without being much considered its integrity checking.
We introduce the problem of 
detecting a corrupted area in a 2-dimensional space, and
investigate two possible efficient approaches
and show their time and space complexities. 
Also, we briefly introduce the idea of an approximation scheme using
a hash sieve and suggest a novel ``adaptive tree'' structure
revealing granularity of information.

\end{abstract}

\section{Introduction}
In recent years, with the help of the development of fast and inexpensive computers, 
many emerging applications have adopted 2-dimensional data such as image or GIS
as well as 1-dimensional data to provide various and advanced services to customers.   
However, such 2-dimensional data can be partially changed or modified (maliciously or accidentally).

In 2007, ``Mayo Clinic and IBM have created a collaborative research facility aimed at 
advancing medical imaging technologies to improve the quality of patient care 
by providing quicker critical diagnosis, such as the growth or shrinkage of tumors \cite{IBM07}''.
There are a few related work such as authentication and recovery image  
by using a digital watermarking method \cite{ZaMu07}.
However, their models are quite different from our model. 

The main difference is that they actually modify an original data 
by adding watermarking information to the original data to detect the modification of image, 
and they do not guarantee the detection of tampering of data 
by allowing a small changes of bits of image without being detected.
Since they do not use (signed) hash values, they do not require to store additional data structure
for detecting data modification.

Moreover, \cite{ZaMu07} only consider pure image data 
rather than general 2-dimensional data (e.g., GIS as well as images). 
Motivated by this, we investigate the problem of efficiently finding a corrupted region
in a given 2-dimensional space.

The rest of the paper is organized as follows. 
We present models and assumptions in Section \ref{ModelsAndAssumptions}.
Then we investigate two approaches in Section \ref{Model1}, 
and Section \ref{Model2}.
We also briefly introduce other possible approaches in Section \ref{OtherApproaches},
and conclude in Section \ref{Conclusion}.

\section{Models and Assumptions}
\label{ModelsAndAssumptions}
Given a 2-dimensional space $N = m \times m$, we want to find a corrupted area $C$
which is composed of $t$ number of times of $c$ where $c$ is the smallest unit (e.g., a pixel).
Suppose Alice has the original data of space $N$.
Bob has an actual data of space $N$ which might be corrupted,
and he wants to detect a corrupted area $C$ if $N$ is corrupted 

We assume that a corrupted area is a single connected shape.
Since there is no prior information in a given 2-dimensional area,
we do not know whether the area was corrupted or not, and   
do not know the size or shape of the corrupted area.
Also, we assume that all data in a given region are uniformly distributed.
That means that all data have the same level of importance to be scanned 
and thus have the same granularity of importance.

The heart of the problem
is finding one corrupted cell, because once we
know such a cell we can ``spread out'' from that cell 
to the remaining corrupted one in breadth-first like fashion
and in time $O(C)$.
This is why we focus in this section on the problem of
finding one corrupted cell.

We consider two models in this paper. 
Suppose Alice's original data and Bob's actual data are stored in a same computer.
Then two sets of data can be compared directly without storing any additional data structure
using a probabilistic method and we will describe it in Section \ref{Model1}.
Note that this model requires an assumption that Alice's original data have not been altered, 
and should be kept in a safe storage.

However, it is hard to achieve such safe devices in our general computing environment.
Computers and their storages are vulnerable to many types of malicious behaviors
such as malware infection, computer bug, or hacking.    
In addition, if Alice's original data and Bob's actual data are stored in different machines, 
then direct comparison is not appropriate. 

Thus, in this model we use signed hash values, assuming that Bob already downloaded these signed hash values 
and its verification key from a server or ISP. Note that the signature key is not stored.
We will describe this second model in Section \ref{Model2}. 
Note that a server could be trusted signing server 
or untrusted distributor of signed hash values proposed in our previous paper \cite{AtChKu08}.

\section{Finding a Corrupted Cell Using a Probabilistic Method}
\label{Model1}

\begin{figure*}
\centerline{\subfigure[A small yellow rectangle is a corrupted area]{\includegraphics[width=.33\textwidth,height=.20\textheight]{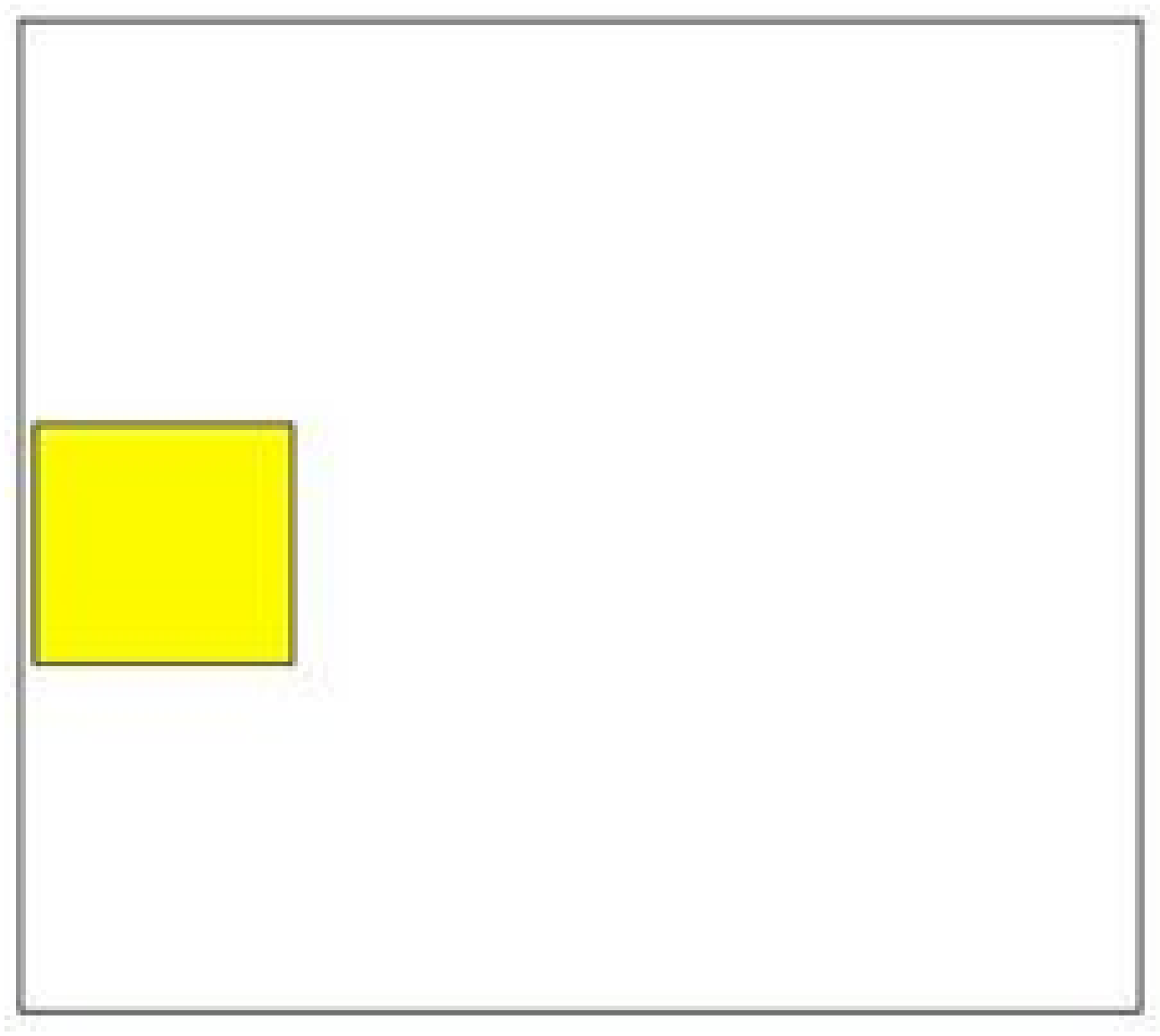}
\label{fig1}}
\hfil
\subfigure[Tiny blue rectangles are randomly chosen cells in each step until find a tiny red corrupted cell in a corrupted area]{\includegraphics[width=.33\textwidth,height=.20\textheight]{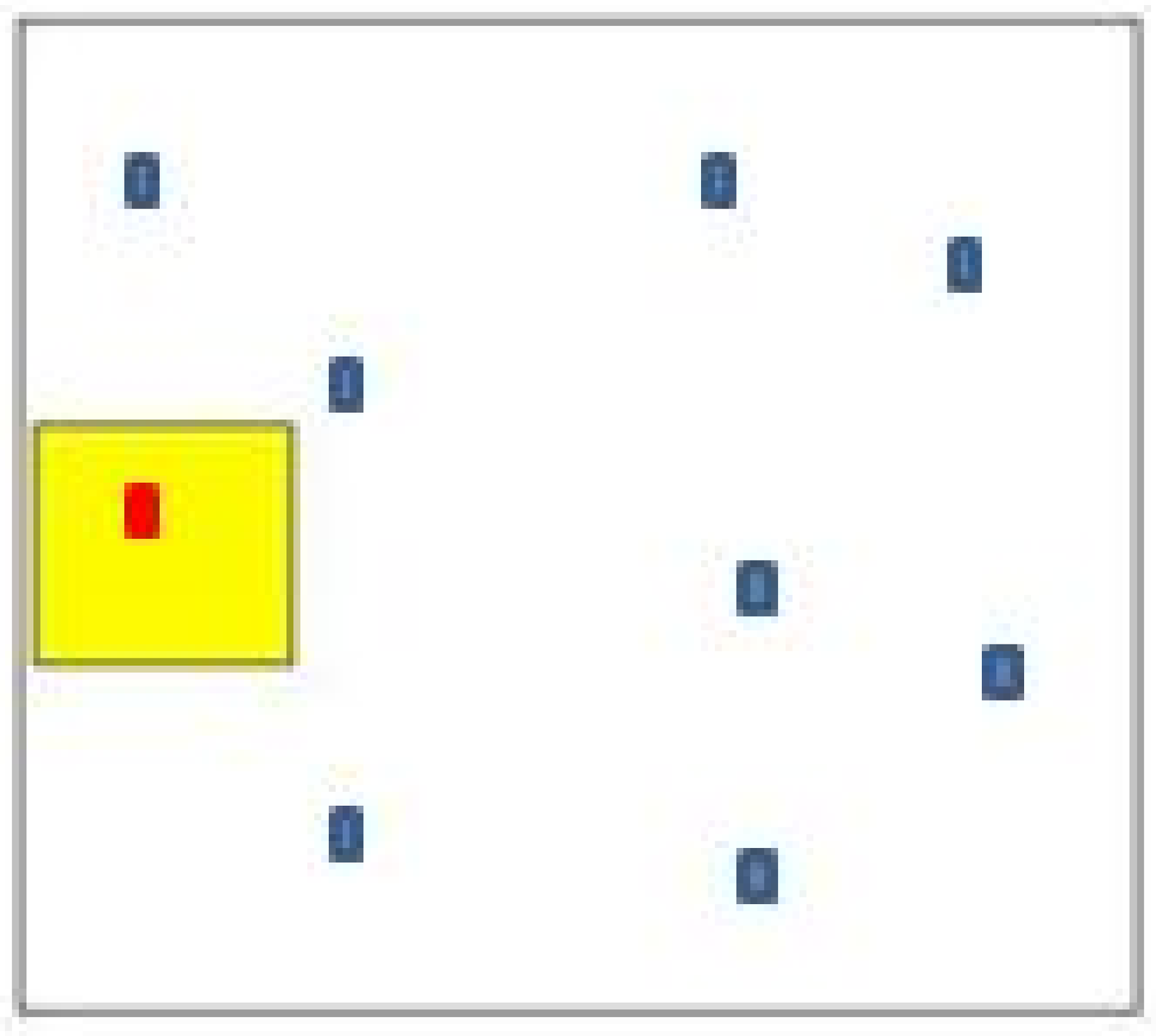}
\label{fig12}}}
\caption{Probabilistic Method}
\label{fig_fig1}
\end{figure*}

Let the random variable $X$ be the number of trials needed to obtain a success to find $c$,
and let $P\{X=k\}$ be the probability of success to find $c$ (Fig. \ref{fig_fig1}).
If $k \leq N-C$ then 
$P\{X=k\} = \bigl(1 - \frac{C}{N}\bigr)\bigl(1 - \frac{C}{N-1}\bigr)\bigl(1 - \frac{C}{N-2}\bigr) \cdots \bigl(1 - \frac{C}{N-(k-2)}\bigr) \bigl(\frac{C}{N-(k-1)}\bigr)$. 
Note that if $k > N-C$ then $P \{X=k\}= 1$.
Let $E[X]$ be the expected number of trials to find $c$ at $k^{th}$ trial. Then 
\begin{eqnarray*}
E[X] &=& \sum_{k=1}^{N} kP\{X=k\} 		\\
	  &=& \biggl(1 - \frac{C}{N}\biggr) + 2\biggl(1 - \frac{C}{N-1}\biggr) + \cdots + (k-1)\biggl( {1 - \frac{C}{N-(k-2)}} \biggr)  + k\biggl(\frac{C}{N-(k-1)}\biggr)
\end{eqnarray*}
If we assume that $C \ll N$ then $\frac{C}{N} \approx \frac{C}{N-1} \approx \cdots \approx \frac{C}{N-(k-1)}$ 
is $p$ where $p = \frac{C}{N}$, and $N-C$ is approximate to $N$. 
Then we have an approximation of $E[X]$ as follows :
\[E[X] = \sum_{k=1}^{\infty} k(1-p)^{k-1}p = \frac{1}{p} = \frac{N}{C}\]
Thus the time to find $c$ is $\frac{N}{C}$. 
Once we find $c$, we can find $C$ by spreading up to $4t$.
Therefore, the total time to find a corrupted area $C$ is $O(\frac{N}{C} + t)$.
Note that this scheme does not require to store any additional data structure to find a corrupted area $C$.

\section{Detecting a Corrupted Cell Using Signed Hashes}
\label{Model2}

We consider two cases -- one in which the time is measured
by the number of cells whose hashes are computed, and the other in which
time is measured by the number of signature verifications.
The latter is motivated by the fact that a signature verification is 
much more time consuming than a cryptographic hash computation.

\subsection{Minimizing Signature Verifications}
\label{SignedHashes}
In this model we store a number $s(N)$ of signatures for 
some subsets of the $m \times m$ cells, hence $s(N)$
is the space complexity of the scheme.   The signature
key is not locally stored to protect against compromise of 
the local machine (that would compromise the signature key).
In the search for
a corrupted cell, a basic comparison operation consists of
checking whether the signature for a subset of the $N$ cells
matches the current values at those cells; such a comparison
costs 1 in the model considered here (even though the comparison
involves looking at all the cells of the subset, it counts
as 1 because there is only 1 signature verification that it
does).

\subsection{A preliminary Scheme}
\label{PreliminaryScheme}

\begin{figure*}
\centerline{\subfigure[Recursive Quadrants]{\includegraphics[width=.37\textwidth,height=.25\textheight]{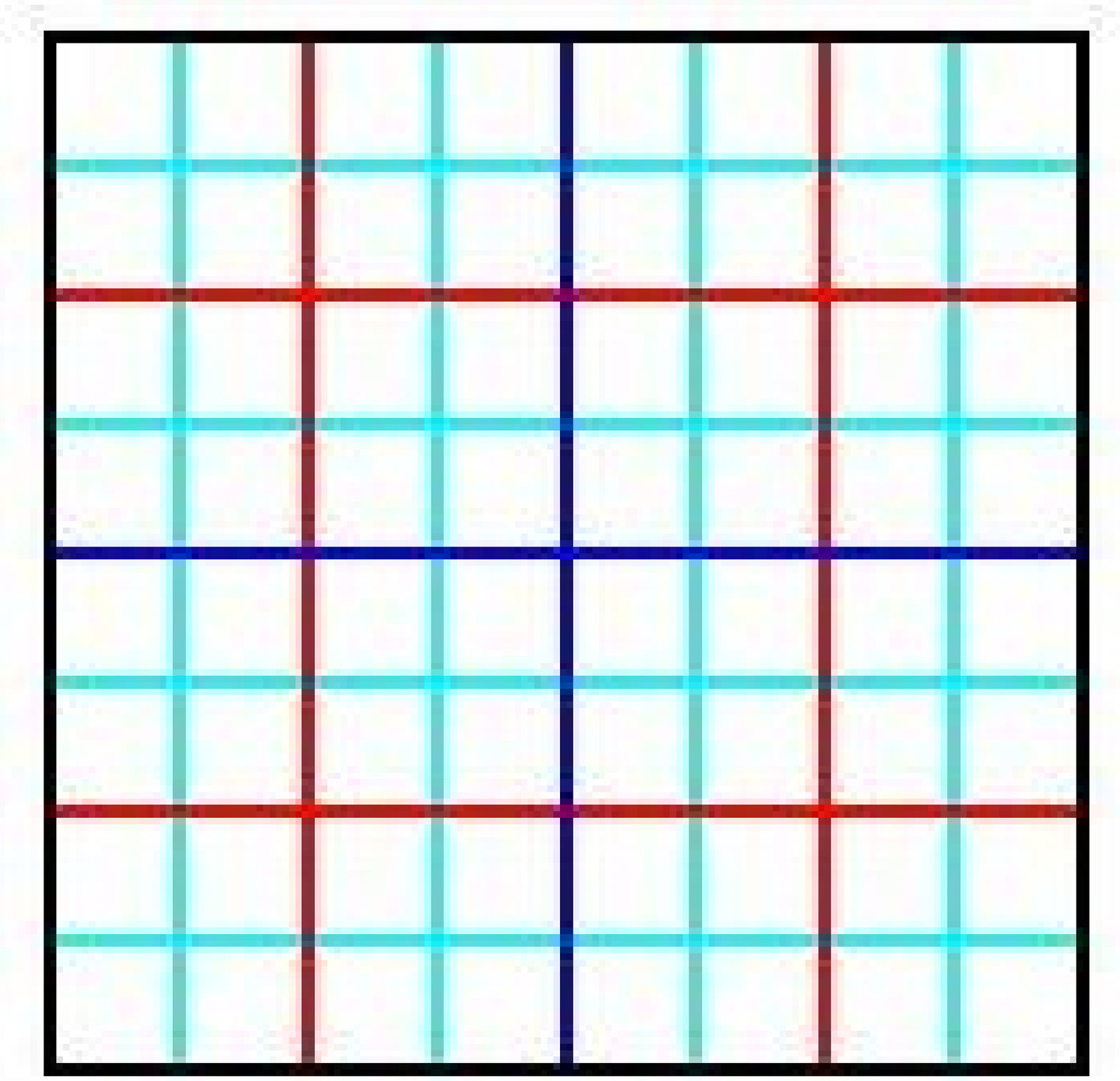}
\label{fig_quadtree1_a}}
\hfil
\subfigure[Signed Hash Values Corresponding Recursive Quadrants]{\includegraphics[width=0.9\textwidth,height=.25\textheight]{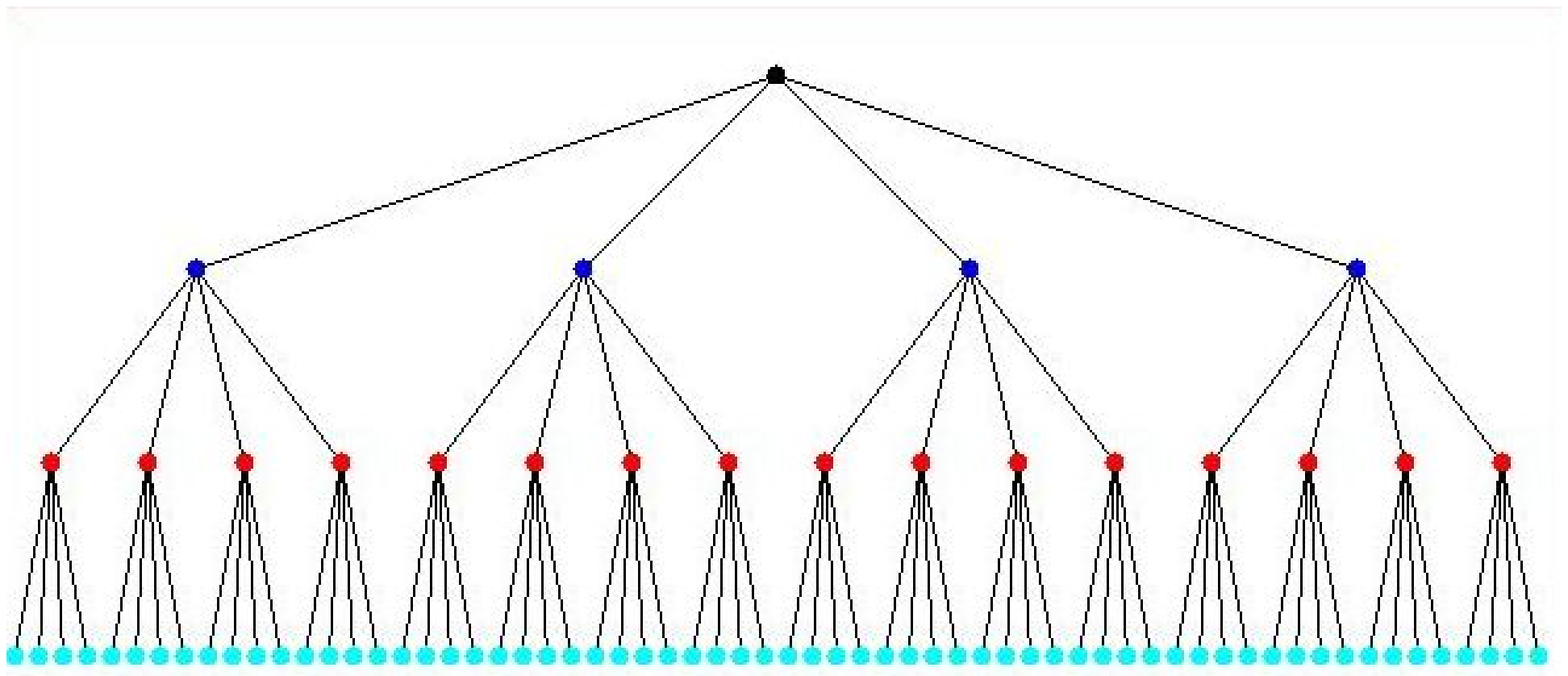}
\label{fig_quadtree1_b}}}
\caption{Preliminary Method}
\label{fig_quadtree1}
\end{figure*}

We store a signed hash for each of the four $(m/2) \times (m/2)$ 
quadrants, and recursively so on for each quadrant (Fig. \ref{fig_quadtree1}).  

Thus the
space complexity obeys the recurrence

$$s(N) = 4 + 4 S(N/4)$$

$$S(1) = 1$$

\noindent This implies $S(N) \leq c_1 N - c_2$,
where $c_1 , c_2$ are constant and it can be proved by induction
as follows.  The basis of the induction is $N=1$,
in which case we do have $1 \leq c_1 * 1 - c_2$ as
long as we choose $c_1 \geq 1 + c_2$. For
the inductive step, we assume the claim holds
for values smaller than $N$ and we show that it
must then hold for $N$.  Using the induction
hypothesis in the recurrence equation gives:

$$s(N) = 4 + 4 S(N/4) \leq 4 + 4 (c_1 N / 4 )  - 4c_2
=4 + c_1 N - 4 c_2$$

\noindent which is $\leq c_1 N - c_2$ as long as we choose
$c_2$ such that $4 -3c_2 \leq 0$, i.e., $c_2 \geq 4/3$.

This structure allows searching for a corrupted cell in
logarithmic time, as follows:
Check if each of the 4 quadrants matches the corresponding
stored signature, and if not recurse on one of
the quadrants that
does not match its stored signature (if more than 
one does not match then recurse on any one 
that does not match).

The number $T(N)$ of signature verifications done
obeys the recurrence

$$T(N) \leq 4 + T(N/4)$$

$$T(1) = 1$$

\noindent and therefore $T(N)$ is $O( \log N )$.

\subsection{An Improved Scheme}
\label{ImprovedScheme}

The following scheme can substantially outperform
it if $C$ is large.

We first show that, in a 1-dimensional version of
the problem (i.e., an array of $N$ entries of which
$C$ are corrupted), we can find a corrupted cell
in a binary search fashion by doing
$t = \log ( N / C )$ signature verifications.
This is done as follows:  First we test the left half of
the array, then the right half, and if both are
corrupted then we have already
found a corrupted cell (the middle cell) otherwise
we recurse in one of the two halves (the
corrupted one).   After $i$ such 
iterations the search region has shrunk down 
to the size of $N / 2^i$, and success at finding
a corrupted cell must occur prior to 
reaching a situation where $N / 2^i > C$,
which is equivalent to saying that 
$ i < \log ( N / C )$.   

To use a 1-dimensional approach as part of
a solution of a 2-dimensional problem, we need
to assume that the $C$ damaged cells form a region 
that is not only connected but also convex
in the sense that for any row or column the 
subset of corrupted cells on that row or 
column forms a connected 1-dimensional region.

The above 1-dimensional solution suggests the 
following approach to the 2-dimensional problem:
Modify the 2-dimensional preliminary scheme
given in the previous subsection 
so that it operates in a manner
similar to the preliminary scheme except that 
we would switch to (one or two) 1-dimensional problem(s)
as soon as we reach a situation where more than 
one quadrant is found to be corrupted: 

\begin{figure*}
\centerline{\subfigure[``$+$'' Boundary]{\includegraphics[width=.30\textwidth,height=.20\textheight]{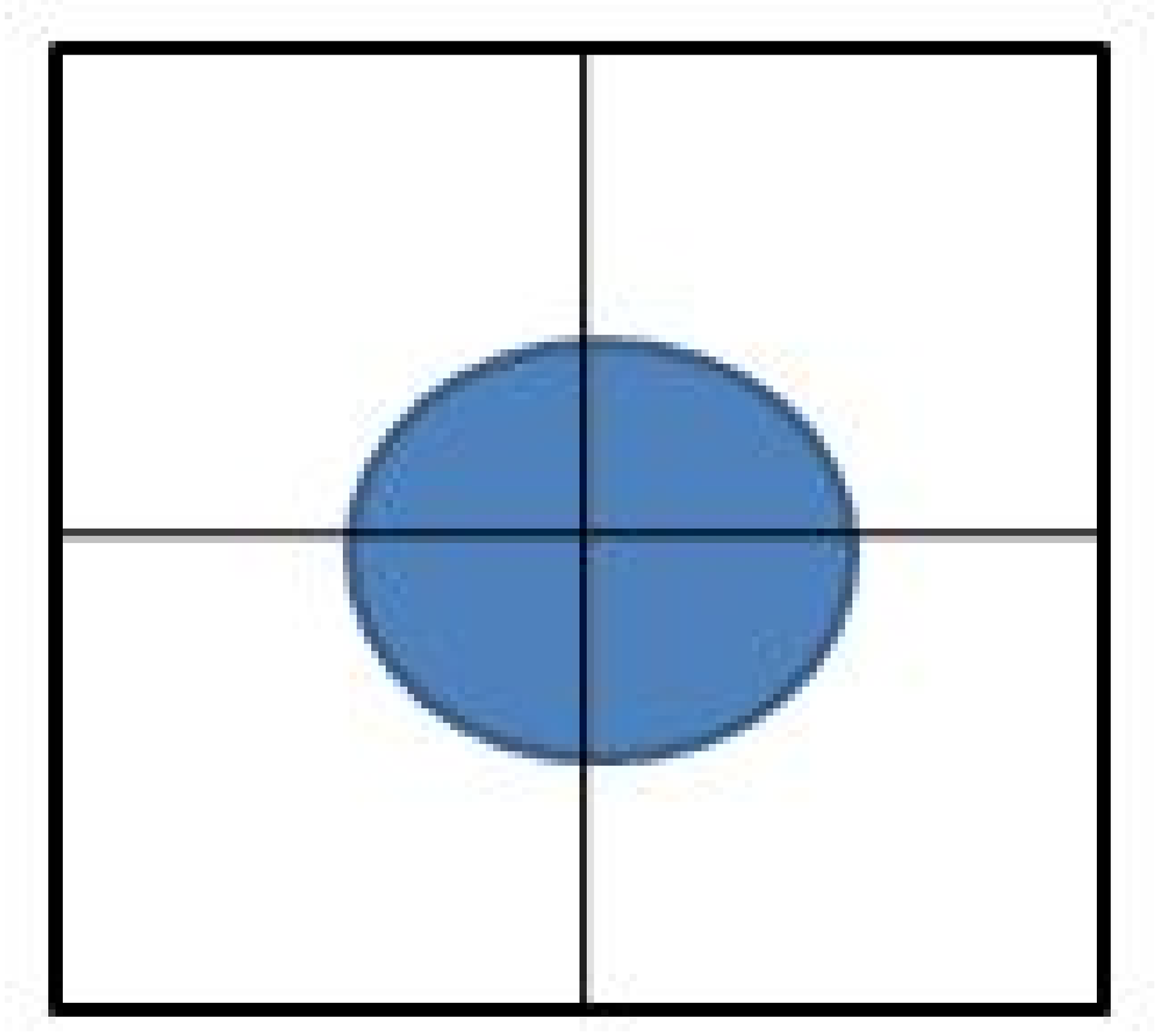}
\label{fig_improve_a}}
\hfil
\subfigure[``T'' Boundary]{\includegraphics[width=.30\textwidth,height=.20\textheight]{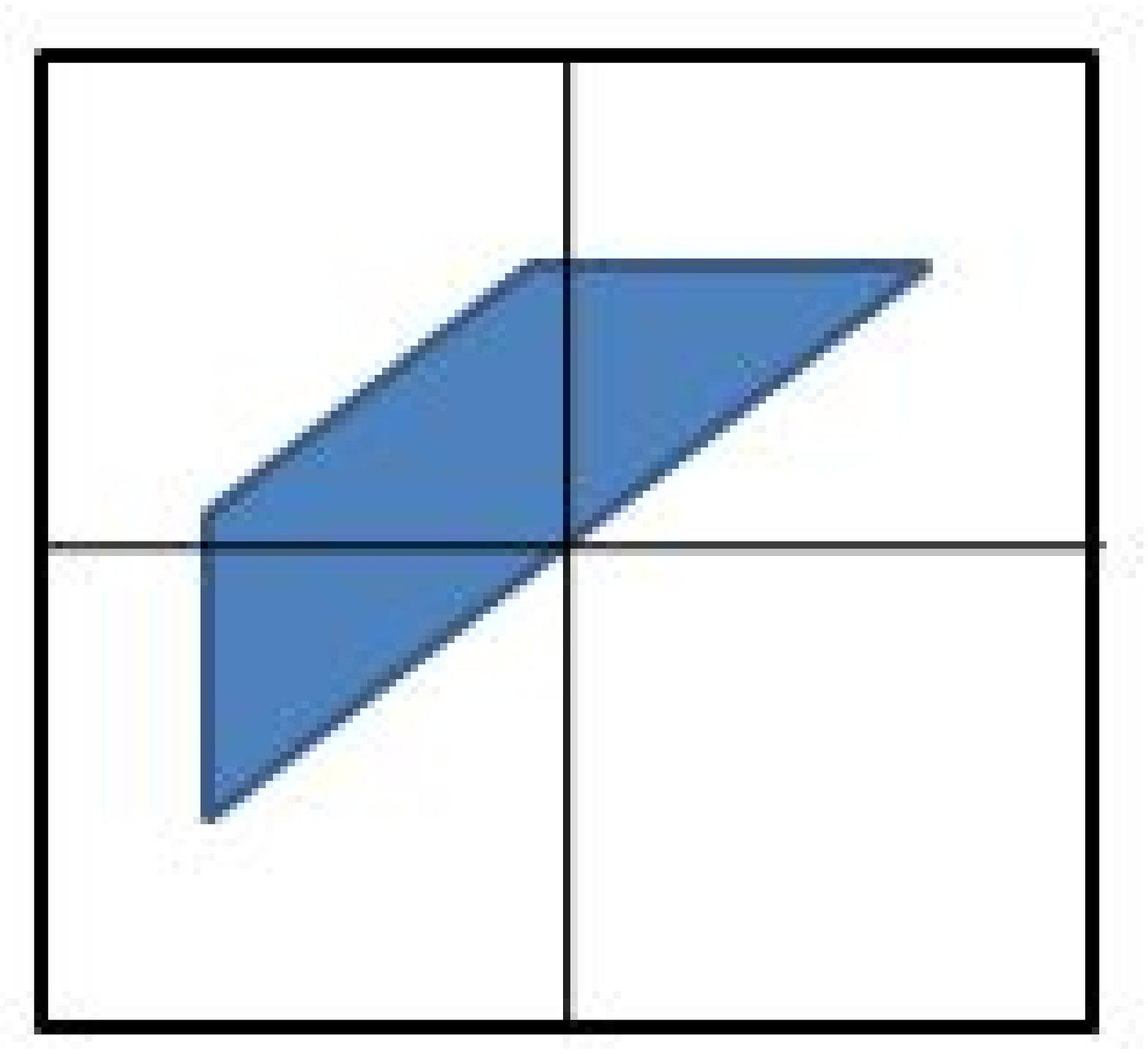}
\label{fig_improve_b}}
\hfil
\subfigure[``I'' Boundary]{\includegraphics[width=.30\textwidth,height=.20\textheight]{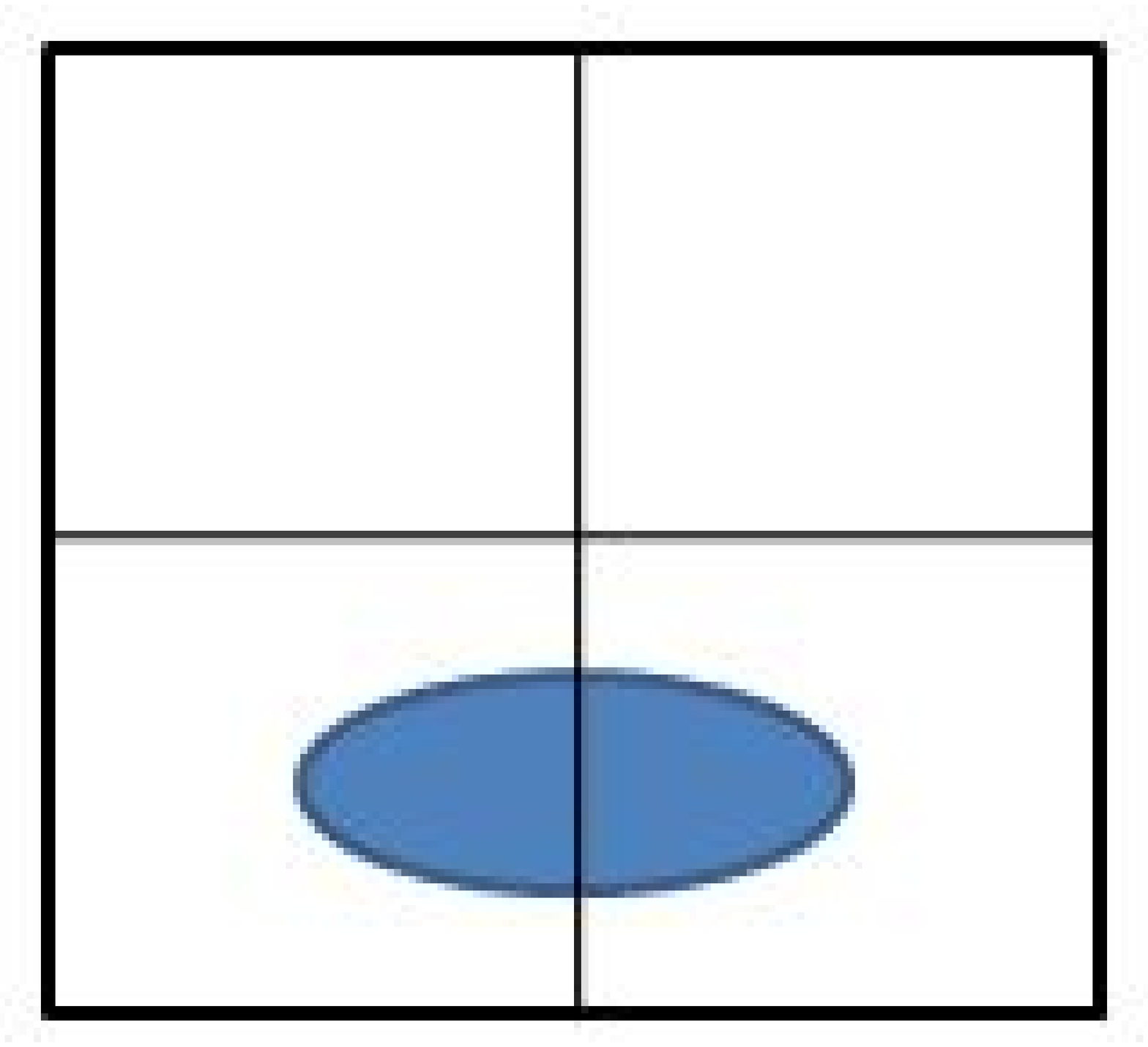}
\label{fig_improve_c}}}
\caption{Improved Scheme}
\label{fig_quadtree}
\end{figure*}

We would then switch to the union of the 
boundaries between all pairs if corrupted 
quadrants.  This union could be shaped like 
a cross (``$+$''), a ``T'', 
an ``I'', a ``--'' , etc (Fig. \ref{fig_quadtree}).  The continuation
of the search would be on the (one or two) one-dimensional
regions that make up that union. Specifically,
if $C'$ of the cells of the resulting 1-dimensional
problems are corrupted, then we would
complete the job in a number of signature
verifications that is $O( \log ( \sqrt{N} / C' ) )$.
The number of signature verifications done
prior to switching to this 1-dimensional mode
is $\leq \log ( N / C )$ because after
$i$ such verifications the size of the square
region being searched has become $N / 2^i$
and we will have surely switched to 1-dimensional
mode before we end up with $N / 2^i < C$.

Of course the above requires storing signatures
for the ``boundary'' one-dimensional problems in
addition to the 4 signatures for the quadrants.
There are at most two such one-dimensional
problems, of size $m = \sqrt{N}$ each.
Because the number of signatures for a one-dimensional 
problem of size $\sqrt{N}$ is $2 \sqrt{N} - 1$,
the space complexity now obeys the recurrence

$$s(N) = 4 + 2 ( 2 \sqrt{N} -1 ) + 4 S(N/4) = 
2 + 4 \sqrt{N} + 4 S(N/4)$$

$$S(1) = 1$$

\noindent 
We show now that the above implies $S(N) \leq c_1 N - c_2 \sqrt{N}$,
where $c_1 , c_2$ are constant (in other
words the space is still $O(N)$, no worse
than in the preliminary scheme).  The
proof, by induction on $N$, is
as follows.  The basis of the induction is $N=1$,
in which case we do have $1 \leq c_1 - c_2$ as
long as we choose $c_1 \geq 1 + c_2$. For
the inductive step, we assume the claim holds
for values smaller than $N$ and we show that it
must then hold for $N$.  Using the induction
hypothesis in the recurrence equation gives:

$$s(N) = 2 + 4 \sqrt{N} + 4 S(N/4) \leq 2 + 
4 \sqrt{N} + 4 (c_1 N / 4 )  - 4 c_2 \sqrt{N/4}
= 2 + c_1 N - 2 c_2 \sqrt{N}$$

\noindent which is $\leq c_1 N - c_2 \sqrt{N}$ as 
long as $2 - c_2 \sqrt{N} \leq 0$, 
i.e., $c_2 \geq 2 / \sqrt{N}$, which
is true as long as we choose $c_2 \geq 2$.

\subsection{Analysis of Preliminary Scheme}
\label{PreliminaryAnalysis}
This section briefly analyzes the scheme of the previous section.

The preliminary scheme's time recurrence is

$$T(N) = N + T(N/4)$$

$$T(1) = 1$$

\noindent The solution becomes $T(N) = O(N)$.

The time recurrence of the 1-dimensional scheme  is

$$T(N) = N + T(N/2)$$

\noindent if $N \geq 2C$, and

$$T(N) = N$$

\noindent if $N < 2C$.  The solution is

$$T(N) = N + (N/2) + (N/2^2) + \cdots + (N/2^i)$$

\noindent

where $i$ is the smallest integer for which
$N/2^i < 2C$.  Using
the above equation twice in $T(N) = 2T(N) -T(N)$ gives:

$$T(N) = 2T(N) - T(N) = 2 (N + (N/2) + (N/2^2) + \cdots + (N/2^i))
- (N + (N/2) + (N/2^2) + \cdots + (N/2^i))$$

\noindent which gives $T(N) = 2N - (N/2^i)$.
Since the definition of $i$ implies that 
$C < N/2^i$, we finally get

$$T(N) = 2N - (N/2^i) < 2N - C$$.

The 2-dimensional solution which converted to the 1-dimensional
can be similarly analyzed and its $T(N)$ becomes

$$T(N) < 2N - C + O( \sqrt{N} )$$

\noindent where the lower order term of $\sqrt{N}$ comes
from the 1-dimensional part of the scheme.

The next section provides and analyzes a scheme that is 
specifically deigned for the ``number of cells touched''
cost model.

\section{Minimizing Cells Accessed Using A Sifting Approach}
\label{SiftingApproach}
We now count the cost of verifying the signature for a region $R$ 
to be {\em proportional to the size of that region}. 
That is, it is the number of cells in $R$ (whereas in the previous model we counted it as 1).

We first give a 1-dimensional version of the approach.
For the convenience we assume $N$ and $C$ are powers of 2.
For $k = 1 , 2, 3 , \ldots$ we do the following 
Stage $k$ until we identify a corrupted cell:

\begin{itemize}
\item
Stage $k$ consists of checking the integrity of 
each cell whose position is a multiple of $N/2^k$
and that was not already tested in any of
the earlier stages $1 , 2 , \ldots , k-1$.
\end{itemize}

The number of stages is at most $i$ where
$i$ is the integer for which 
$N/2^i = C$, i.e., i = $\log ( N / C )$.
The work done at stage 1 is 2 (the cells at 
positions $N/2$ and $N$), the work done at stage 2 is 2
(the cells at positions $N/4$ and $3N/4$), at stage 3
it is 4, etc.  In general,
for $k \geq 2$, the work needs to be done at stage $k$ is $2^k - 2^{k-1} = 2^{k-1}$.
The total work is therefore

$$2 + \sum_{k=2}^i 2^{k-1} = O( 2^i )$$

\noindent which is $O( N / C )$.  Note that this
is much better than the $O( N-C )$ achieved by using
the 1-dimensional scheme of the previous section.

\begin{figure*}
\centerline{\subfigure[Stage $k$=1]{\includegraphics[width=.25\textwidth,height=.15\textheight]{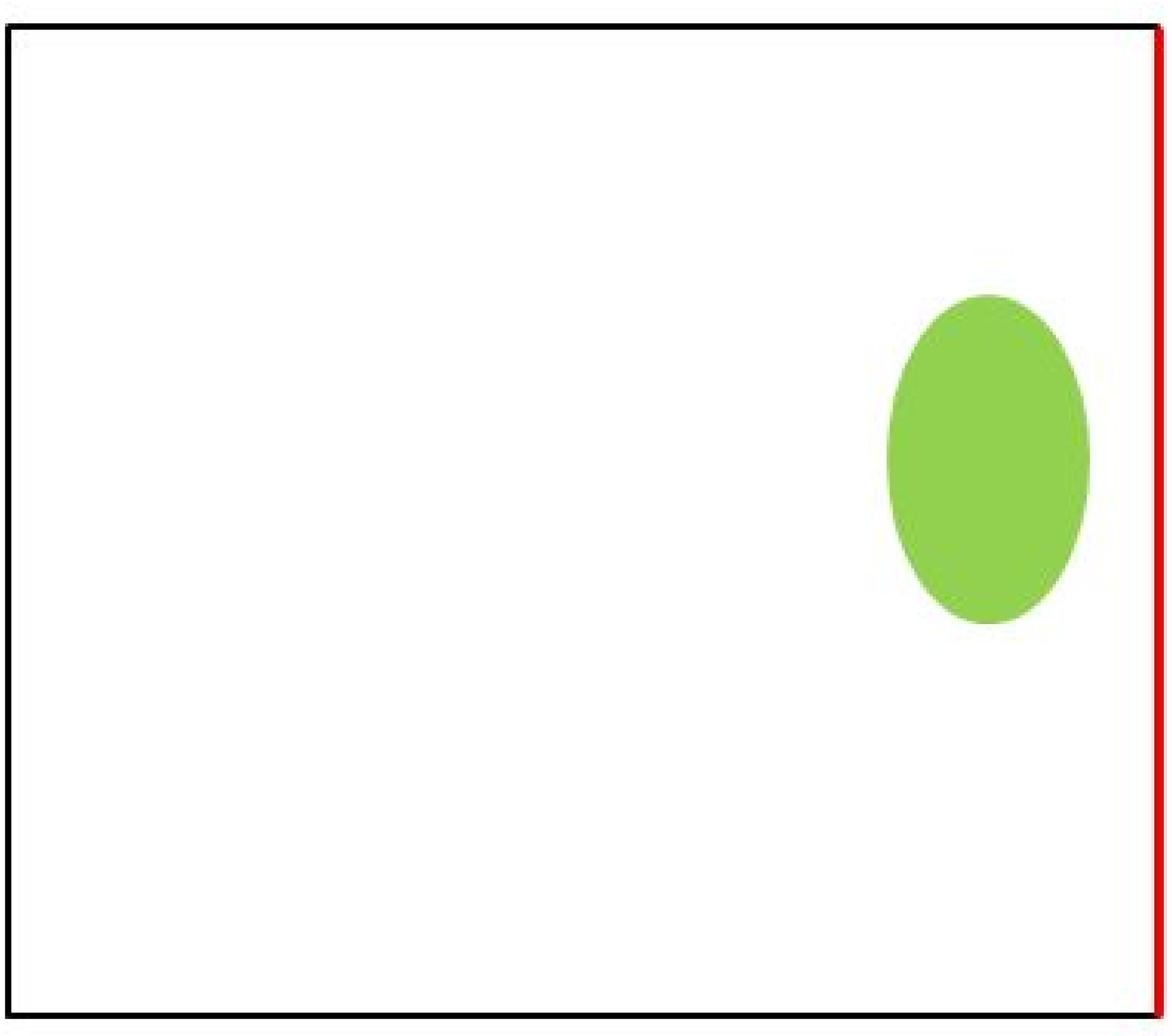}
\label{fig_1}}
\hfil
\subfigure[Stage $k$=2]{\includegraphics[width=.25\textwidth,height=.15\textheight]{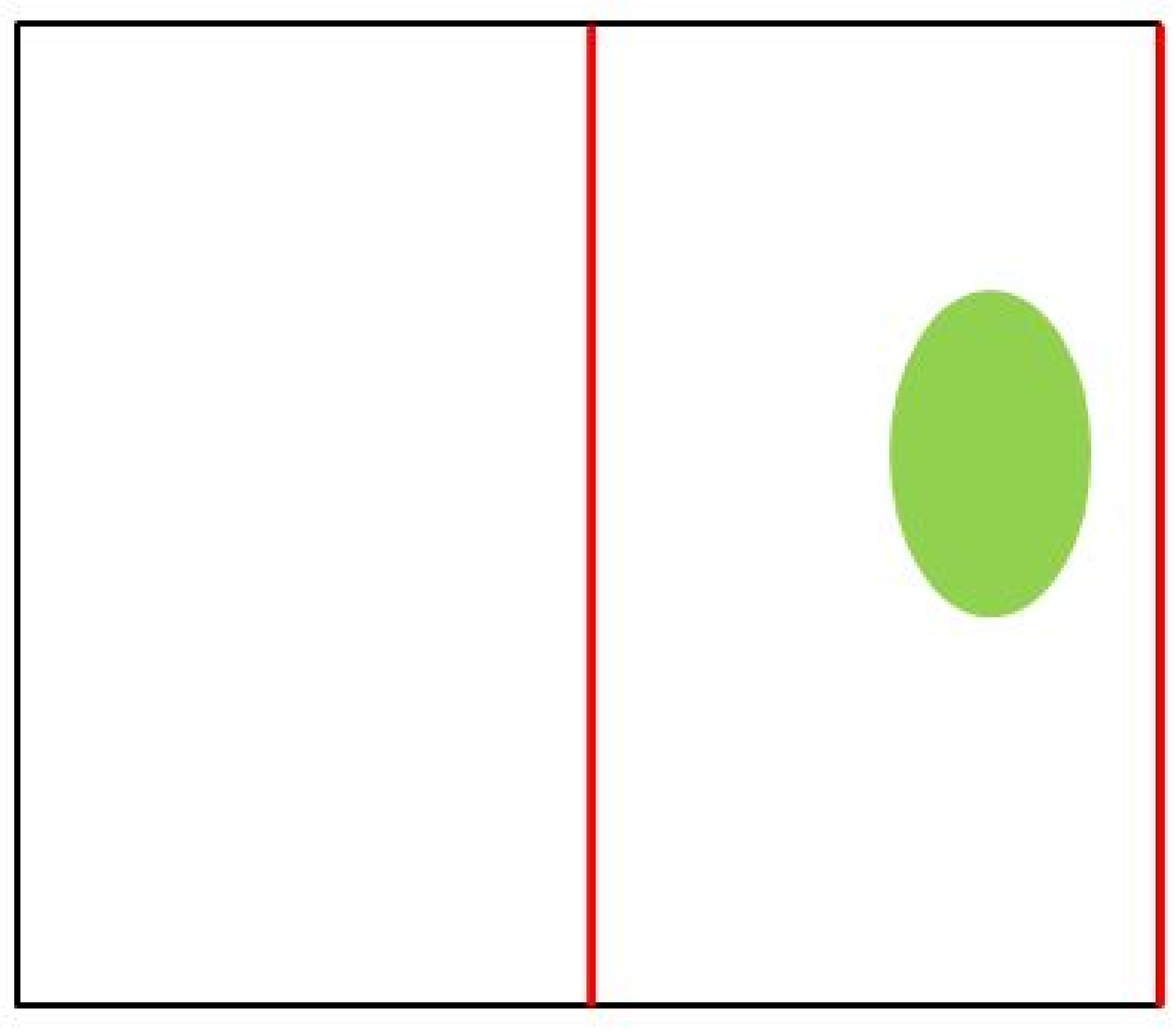}
\label{fig_2}}
\hfil
\subfigure[Stage $k$=3]{\includegraphics[width=.25\textwidth,height=.15\textheight]{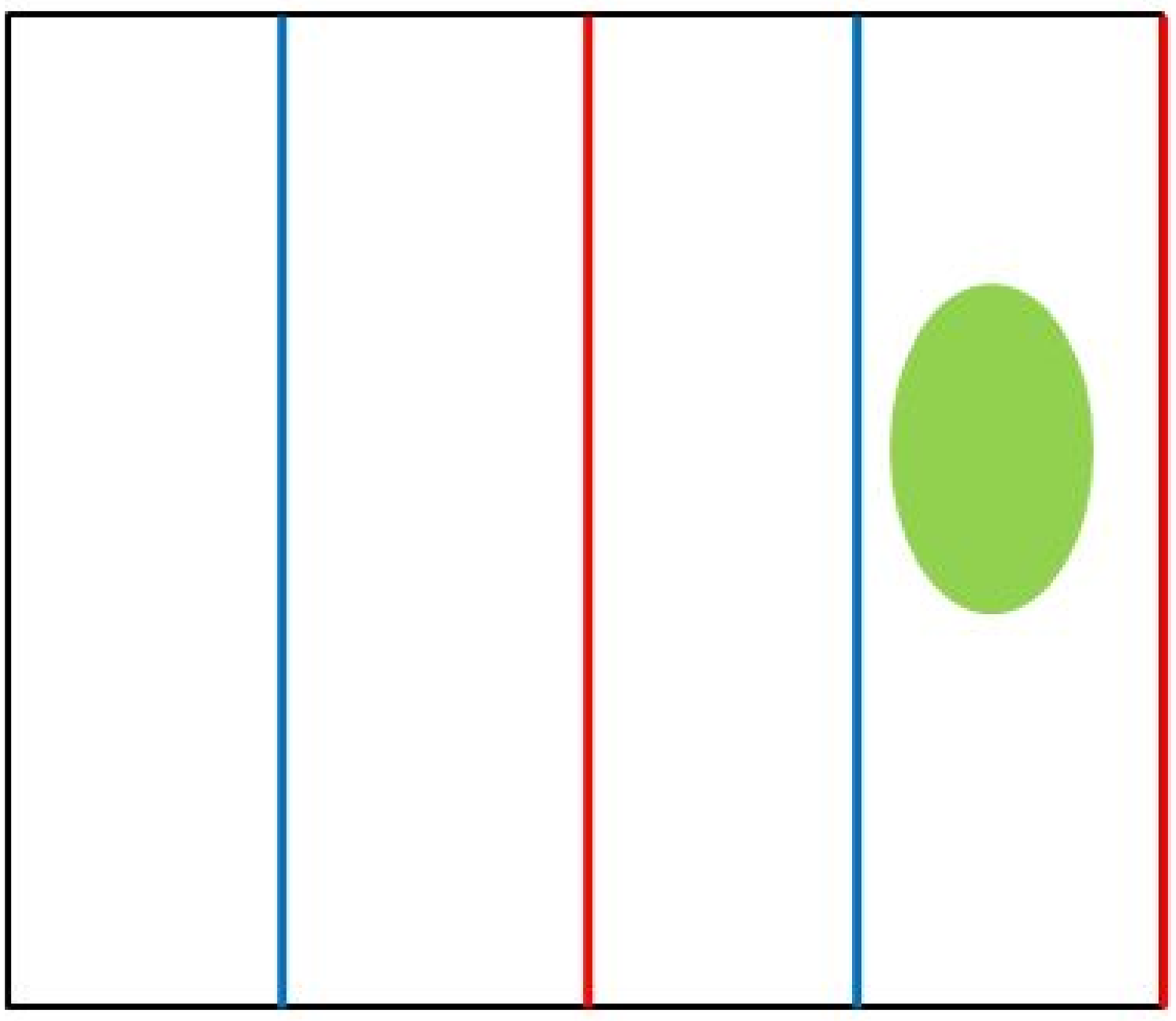}
\label{fig_3}}
\hfil
\subfigure[A corrupted oval green area is detected at stage $k$=4 ]{\includegraphics[width=.25\textwidth,height=.15\textheight]{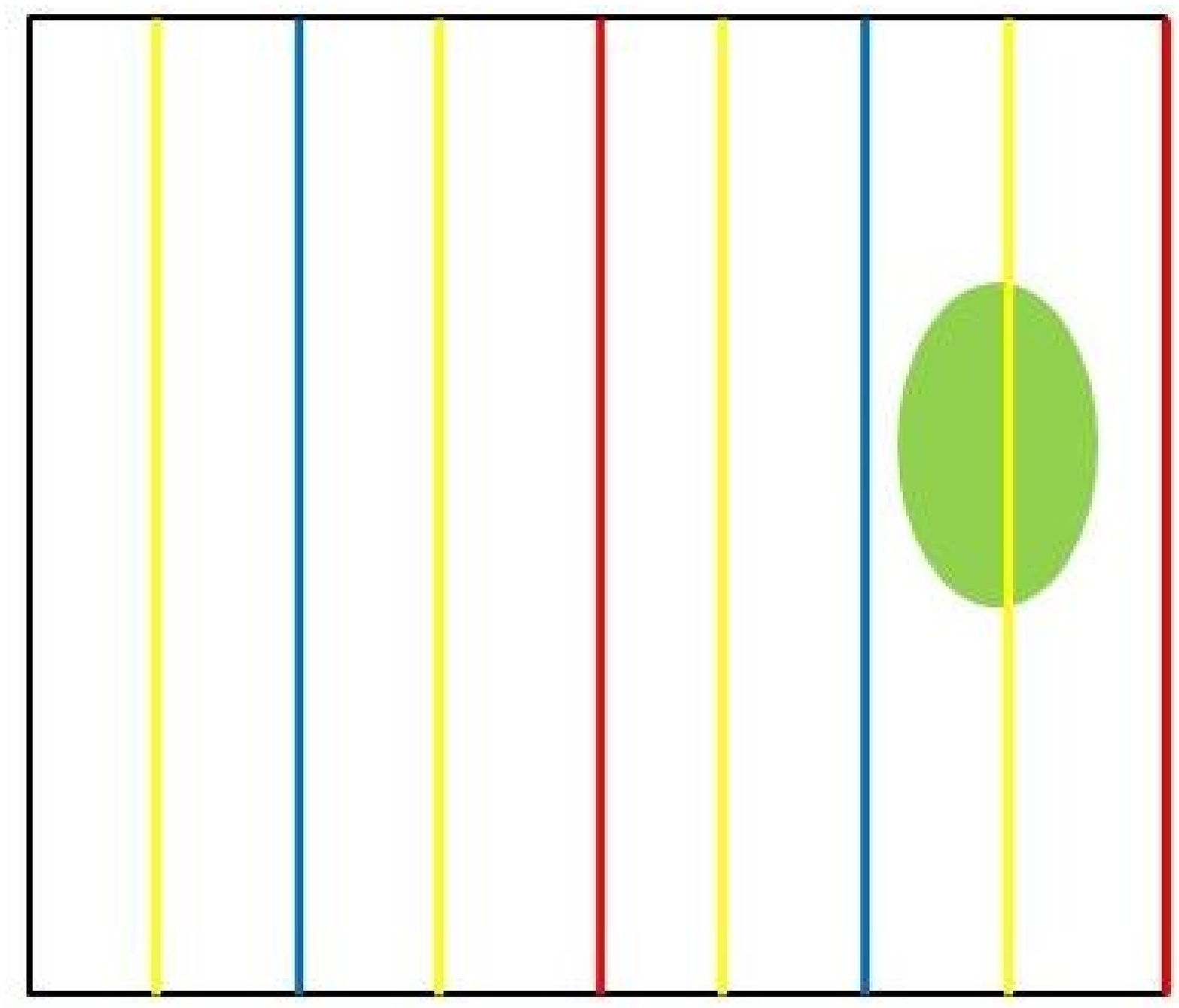}
\label{fig_5}}}
\caption{A Sifting Approach}
\label{fig_shifting}
\end{figure*}

The scheme extends to 2 dimensions by using it
on the columns and then, once a column with corrupted
cells has been found, using the above 1-dimensional 
on that column.  The process of
identifying a corrupted column is as follows (Fig. \ref{fig_shifting}). 
For $k = 1 , 2, 3 , \ldots$ we do the following
Stage $k$ until we have identified a corrupted cell:

\begin{itemize}
\item
Stage $k$ consists of checking the integrity of
each column whose number is a multiple of $N/2^k$
{\em and} that was not already tested in any of
the earlier stages $1 , 2 , \ldots , k-1$.
\end{itemize}

The number of stages is at most $i$ where
$i$ is the integer for which
$N/2^i = C$, i.e., i = $\log ( N / C )$.
The work done at stage 1 is $2 \sqrt{N}$ (the columns 
numbered $N/2$ and $N$), the work done at stage 2 is $2 \sqrt{N}$
(the columns at positions $N/4$ and $3N/4$), at stage 3
it is $4 \sqrt{N}$, etc.  In general,
for $k \geq 2$, the work done at stage $k$ is 

$$( 2^k - 2^{k-1} ) \sqrt{N} = 2^{k-1} \sqrt{N}$$

The total work is therefore

$$ ( 2 + \sum_{k=2}^i 2^{k-1} ) \sqrt{N} = O( 2^i \sqrt{N} )$$

\noindent which is $O ( N^{1.5} / C )$.  
Once a column with corrupted cells is identified, 
we can find a corrupted cell within that column
in time $O( \sqrt{N} )$, so the dominant cost
is the $O ( N^{1.5} / C )$ for finding the column.

\subsection{Hybrid Approach}
\label{HybridApproach}
Let $A_1$ be the improved scheme and $A_2$ be the sift scheme.
It is clear that if $C < \sqrt{N}$, $A_1$ is better and $A_2$ is better otherwise.
However, we do not know the value of $C$ in advance
since we assume that only information we are given is a 2-dimensional area.
We do not know whether it is corrupted or not. Neither do we know the size of corrupted area.

However, we can perform the previous one ``in parallel'' 
by alternating between $A_1$ and $A_2$. We stop as soon as one of them succeeds.
We do this because we do not know the value of $C$ in advance.
Thus we cannot choose which of the two is better for that unknown $C$.  
The overall complexity of this hybrid solution for the 2-dimensional case becomes:

$$T(N) = O( N * \min \{ 1 , \sqrt{N}/C \} )$$

\section{Other Approaches}
\label{OtherApproaches}
In the following sections, we briefly introduce two different approaches\footnote{We will provide a detail description of these approaches in our full version of this paper very soon.} by relaxing assumptions that we used in previous sections.
One is an approximation scheme, and the other is an adaptive tree structure.

\subsection{Approximation Scheme}
First, we give an idea of an approximation scheme 
by reducing the number of signed hash values 
using a sparse spacing filling curve (e.g, Hilbert curve or Z-curve) \cite{LiJe00,OrMa98}.
The key idea is that we store two layers of hash values: horizontal and vertical hash layers,
and each layer consists of $O(\sqrt{N})$ number of hash values of horizontal and vertical layer
respectively such that a sparse space fillings curve of signed hashes will be displayed 
when those two layers are overlapped.
In this case, the time to find a corrupted cell $T(N)$ will be $O(\sqrt{N})$, 
and the size of stored hash values $s(N)$ as well which are much better. 

Since the overlapped layers form a sieve so that we can sieve a corrupted area,
we can find an approximate corrupted area $C'$ 
(instead of a precise shape of corrupted area $C$ 
without spreading after we found a corrupted cell),
and this might help to detect multiple corrupted areas.

However, one of issues in this model is the level of sparseness of a space filling curve.
Even though a space filling curve reduces a dimension by one, it is quite dense. 
Thus we want to use a sparse version, 
but it is not easy to measure how much we should make it sparse 
since it is closely related with the size of corrupted area which is not a given information in advance.

\subsection{Adaptive Tree}

In this section, we consider the case when data do not have the same level of importance.
For example, an image of internal organs of human body probably has 
different probabilities of having tumors; 
The probability of having tumors in layers of fat or water will be negligible
unlike lung or stomach.  
In this model, we can assign different levels of granularity of importance in a region.

\begin{figure*}[htp]
\centering
{\includegraphics[width=.95\textwidth,height=.3\textheight]{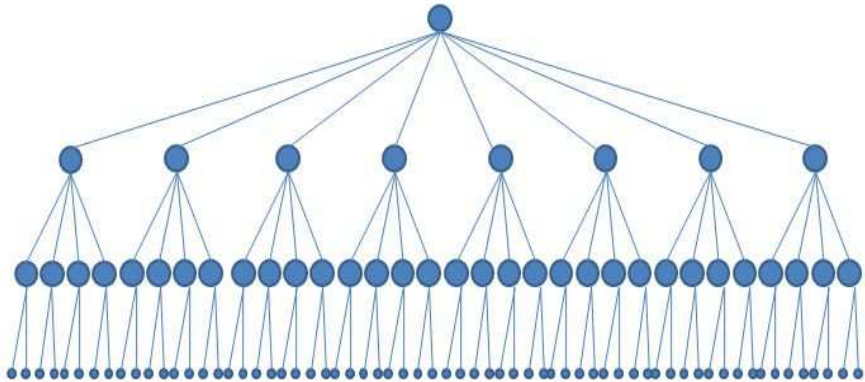}}
\caption{Adaptive Tree (e.g.,$N = 64$)}
\label{fig_AdaptiveTree}
\end{figure*}

Motivated by this, we introduce a novel tree structure, ``adaptive tree'', 
for signed hash values which includes granularity information.
We create a tree such that the number of children of each node at $h$ is $2^h$ 
where $h$ is a height of a tree.
The degree of a node is $2^h$ at depth 0 (i.e. root) and $2^{t-1}$ at depth 1 and then 
$2^1$ at depth $h-1$. Thus $$2^h 2^{h-1} 2^{h-2} \cdots 2^1 = N$$ and 
$h \approx  \sqrt{2 \lceil (log N + 1 ) \rceil }$ (Fig. \ref{fig_AdaptiveTree}).

Thus the height of a tree $h$ is $O( \sqrt{log N} )$.
Note that the size of each child becomes $N / (2^{\sqrt{log N}})$. 
This adaptive tree structure has a shorter height, 
and it gives you the more detail information (i.e., smaller granularity),
the more you go down the tree.

\section{Conclusion}
\label{Conclusion}

We introduced the problem of detecting a corrupted area in a 2-dimensional space,
and investigated some possible solutions under the two different models efficiently, 
and analyzed their time and space complexities.
Our approaches do not tolerate even 1-bit modification of original data unlike watermarking schemes \cite{ZaMu07}.
We also briefly described approximate scheme with much better time and space complexity by relaxing assumptions.
Furthermore, we introduce a novel ``adaptive tree'' structure revealing granularity of information.
We will provide a full extended version of this paper in a very near future.

\bibliographystyle{abbrv}
\bibliography{DetectCArea}
\end{document}